\documentclass[aps,prl,twocolumn,groupedaddress,showpacs]{revtex4}

\usepackage{graphicx}
\usepackage{dcolumn}
\usepackage{bm}
\usepackage{amssymb}
\usepackage{amsmath}

\begin{document}

\title{Ring Intermittency in Coupled Chaotic Oscillators at the Boundary of Phase Synchronization}

\author{Alexander~E.~Hramov$^{1}$}
\author{Alexey~A.~Koronovskii$^{1}$}
\author{Maria~K.~Kurovskaya$^{1}$}
\author{S. Boccaletti$^2$}
\affiliation{$^1$ Faculty of Nonlinear Processes, Saratov State
University, Astrakhanskaya, 83, Saratov, 410012, Russia \\
$^2$ CNR --- Istituto dei Sistemi Complessi Via Madonna del Piano,
10 50019 Sesto Fiorentino (FI), Italy}

\date{\today}

\begin{abstract}
A new type of intermittent behavior is described to occur near the
boundary of phase synchronization regime of coupled chaotic
oscillators. This mechanism, called {\it ring intermittency},
arises for sufficiently high initial mismatches in the frequencies
of the two coupled systems. The laws for both the distribution and
the mean length of the laminar phases versus the coupling strength
are analytically deduced. A very good agreement between the
theoretical results and the numerically calculated data is shown.
We discuss how this mechanism is expected to take place in other
relevant physical circumstances.
\end{abstract}

\pacs{05.45.-a, 05.45.Xt, 05.45.Tp}

\maketitle

Intermittent behavior is an ubiquitous phenomenon in nonlinear
science. Its arousal and main statistical properties have been
studied and characterized already since long time ago, and
different types of intermittency have been classified as types
I--III~\cite{Berge:1988_OrderInChaos,Dubois:1983_IntermittencyIII}
or on--off
intermittency~\cite{Platt:1993_intermittency,Heagy:1994_intermittency}.
One of the most general and interesting manifestations of the
intermittent behavior can be observed near of the boundary of
chaotic synchronization regimes. Indeed, close to the threshold
parameter values for which the coupled systems show synchronized
dynamics, it is observed that the de-synchronization mechanism
involves persistent intermittent time intervals during which the
synchronized oscillations are interrupted by the non-synchronous
behavior. These pre-transitional intermittencies have been
described in details for the case of lag synchronization
\cite{Rosenblum:1997_LagSynchro,Boccaletti:2000_IntermitLagSynchro,Zhan:2002_ILS}
and for generalized synchronization
\cite{Hramov:2005_IGS_EuroPhysicsLetters}, and their main
statistical properties (following those of the on-off
intermittency) have been shown to be common to other relevant
physical processes.

As far as intermittency phenomena near the phase synchronization
onset are concerned, two types of intermittent behavior have been
observed so far~\cite{Pikovsky:1997_PhaseSynchro_UPOs,Lee:1998:PhaseJumps,%
Boccaletti:2002_LaserPSTransition_PRL,Rosa:1998_TransToPS}, namely
the type-I intermittency and the super-long laminar behavior (so
called ``eyelet
intermittency''~\cite{Pikovsky:1997_EyeletIntermitt}).

In this Letter we report that a new type of intermittent behavior
is observed near the phase synchronization boundary of two
unidirectionally coupled chaotic oscillators, when the natural
frequencies of the two oscillators  are sufficiently different
from one another. The system under study is represented by a pair
of unidirectionally coupled R\"ossler systems, whose equations
read as
\begin{equation}
\begin{array}{ll}
\dot x_{d}=-\omega_{d}y_{d}-z_{d},& \dot x_{r}=-\omega_{r}y_{r}-z_{r} +\varepsilon(x_{d}-x_{r}),\\
\dot y_{d}=\omega_{d}x_{d}+ay_{d},& \dot y_{r}=\omega_{r}x_{r}+ay_{r},\\
\dot z_{d}=p+z_{d}(x_{d}-c),\mbox{} & \dot z_{r}=p+z_{r}(x_{r}-c),\\
\end{array}
\label{eq:Roesslers}
\end{equation}
where $(x_{d},y_{d},z_{d})$ [$(x_{r},y_{r},z_{r})$] are the
cartesian coordinates of the drive (the response) oscillator, dots
stand for temporal derivatives, and $\varepsilon$ is a parameter
ruling the coupling strength. The other control parameters of Eq.
(\ref{eq:Roesslers}) have been set to $a=0.15$, $p=0.2$, $c=10.0$,
in analogy with previous studies
~\cite{Hramov:2005_GSNature,Harmov:2005_GSOnset_EPL}. The
$\omega_r$--parameter (representing the natural frequency of the
response system) has been selected to be $\omega_r=0.95$; the
analogous parameter for the drive system has been fixed to
$\omega_d=1.0$. For such a choice of parameter values, both
chaotic attractors of the drive and response systems are, at zero
coupling strength, phase coherent. Furthermore, the boundary of
the phase synchronization regime occur around
$\varepsilon_{c}\approx 0.124$.

The instantaneous phase of the chaotic signals $\varphi(t)$ can be
therefore introduced in the traditional way, as the rotation angle
${\varphi_{d,r}=\arctan(y_{d,r}/x_{d,r})}$ on the projection plane
$(x,y)$ of each system. The presence of the phase synchronization
regime can be detected by means of monitoring the time evolution
of the instantaneous phase difference, that has to obey the phase
locking condition~\cite{Boccaletti:2002_SynchroPhysReport}
\begin{equation}
|\Delta\varphi(t)|=|\varphi_{d}(t)-\varphi_{r}(t)|<\mathrm{const}.
\label{eq:PhaseLocking}
\end{equation}
Below the boundary of the phase synchronization regime, the
dynamics of the phase difference $\Delta\varphi(t)$ features time
intervals of phase synchronized motion (laminar phases)
persistently and intermittently interrupted by sudden phase slips
(turbulent phases) during which the value of $|\Delta\varphi(t)|$
jumps up by $2\pi$.

By analyzing the statistics of the laminar phases, it is found
that the intermittent type behavior described in Refs.
~\cite{Pikovsky:1997_PhaseSynchro_UPOs,Lee:1998:PhaseJumps,%
Boccaletti:2002_LaserPSTransition_PRL,Rosa:1998_TransToPS,%
Pikovsky:1997_EyeletIntermitt} takes place only for small
differences in the natural frequencies of the drive and response
systems. In particular, the eyelet intermittent phenomenon occurs
in the range $\omega_d = 0.90 \div 0.98$. As far as large
differences in the natural frequencies of the drive and response
systems are concerned ($\omega_d<0.90$ and $\omega_d>0.98$), a
novel type of intermittent behavior emerges which differs
remarkably from the ones known so far. In the following we will
describe the properties of this new type of behavior, that we
called \textit{ring intermittency}, due to the specific dynamical
mechanism that produces it.

\begin{figure}[tb]
\centerline{\includegraphics*[scale=0.35]{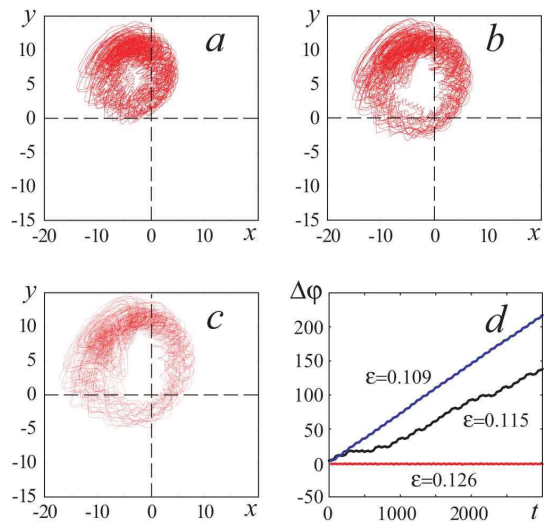}}
\caption{(Color online) The phase trajectory of the response
system on the $(x',y')$-plane rotating around the origin when the
coupling parameter strength is selected as (\textit{a})
$\varepsilon=0.126$
--- the phase synchronization regime, (\textit{b})
$\varepsilon=0.115$
--- the ring intermittency and (\textit{c}) $\varepsilon=0.109$
--- the asynchronous dynamics. (\textit{d}) The dependence of
the phase difference $\Delta\varphi(t)$ on the time $t$ for the
coupling strength values $\varepsilon$ used in
Fig.~(\textit{a--c}).} \label{fgr:RotatingPlane}
\end{figure}

Let us start with discussing the mechanism at the origin of the
arousal of the ring intermittency, and that rules out the scaling
laws characterizing this phenomenon. It is well-known that there
are two different scenarios for synchronization destruction in a
periodic oscillator driven by an external force,  corresponding
respectively to small and large detunings with the external signal
frequency (see, e.g.,
tutorial~\cite{Pikovsky:2000_SynchroReview}). Under certain
conditions (i.e., for the periodically forced weakly nonlinear
isochronous oscillator), the complex amplitude method may be used
to find the solution describing the oscillator behavior in the
form ${u(t)=\mathrm{Re}\,a(t)e^{i\omega t}}$. For the complex
amplitude $a(t)$ one obtains averaged (truncated) equations
${\dot{a}=-i\nu a+a-|a|^2a-ik}$, where ${\nu}$ is the frequency
mismatch, and $k$ is the (renormalized) amplitude of the external
force. For the small $\nu$ and large $k$ the stable solution
${a(t)=Ae^{i\phi}=\mathrm{const}}$ corresponds to the synchronous
regime, with the synchronization destruction corresponding to the
the saddle-node bifurcation on the plane of the complex amplitude.
For large frequency mismatches with the decrease of $k$-value the
fixed point (stable node) on the complex amplitude plane becomes
sequentially a stable focus and an unstable focus (via the
Andronov--Hopf bifurcation). In this case the phase
synchronization destruction is connected with the limit cycle
location on the complex amplitude plane
(see~\cite{Pikovsky:2000_SynchroReview} for detail). When the
limit cycle starts enveloping the origin, the synchronization
regime begins to destroy. Obviously, if one considers the behavior
of the synchronized periodic oscillator on the plane $(x',y')$
rotating with the frequency of the external signal around the
origin he observes the stable node for the small values of the
frequency detuning and a cycle for the large ones, respectively.
These considerations on the rotating plane may be made apparent by
using the coordinate transformation
${x'=x_r\cos\varphi_d+y_r\sin\varphi_d}$,
${y'=-x_r\sin\varphi_d+y_r\cos\varphi_d}$, where
$\varphi_d=\varphi_d(t)$ is the instantaneous phase of the drive
system.

The same effects may be also observed for  chaotic oscillators.
Indeed, in Fig.~\ref{fgr:RotatingPlane},\,\textit{a} the behavior
of the synchronized response oscillator~(\ref{eq:Roesslers}) is
shown on the plane $(x',y')$ rotating around the origin in
accordance with the phase $\varphi_d(t)$ of the drive system when
the control parameters $\omega_d$ and $\omega_r$ are detuned
sufficiently. One can see that the phase trajectory on this plane
looks like a ring. This effect arises insofar as the R\"ossler
system may be considered as a noise smeared periodic oscillator
(see, e.g.,~\cite{Pikovsky:1997PhaseSynchro}). Therefore one
observes \textit{the ring} consisting of the phase trajectories,
instead of the limit cycle that would occur in the periodic case.
It is the case of the large control parameter mismatch that is
accompanied by the ring intermittency behavior, while for the
small parameter detuning the intermittent type-I as well as the
eyelet intermittency are revealed. When the coupling strength
$\varepsilon$ gets below the critical value $\varepsilon_c$ the
phase trajectory on the $(x',y')$-plane starts enveloping the
origin (see Fig.~\ref{fgr:RotatingPlane},\,\textit{b}, the origin
is the point of intersection of the dashed lines), and the phase
synchronization regime begins to destroy, as a phase slip is
observed all the times that the phase trajectory envelops the
origin of that plane. As the coupling strength decreases further,
the phase trajectory envelops the origin more often, and the phase
slips occur more frequently. Finally, when the coupling strength
$\varepsilon$ becomes less than $\varepsilon_t\approx0.1097$ the
origin is inside the ring (see
Fig.~\ref{fgr:RotatingPlane},\,\textit{c}), therefore every
rotation of phase trajectory causes a phase slip. So, varying the
coupling strength $\varepsilon$ one observes: (i) the phase
synchronization regime for ${\varepsilon>\varepsilon_c}$, (ii) the
intermittent behavior for
${\varepsilon_t<\varepsilon<\varepsilon_c}$ and (iii) the
asynchronous dynamics for ${\varepsilon<\varepsilon_t}$ when the
phase slips follow each other at approximately equal time
intervals $T$, the averaged period of the phase trajectory
rotation on the $(x',y')$-plane (see
Fig.~\ref{fgr:RotatingPlane},\,\textit{d}).

Let the probability that the phase trajectory on the rotating
$(x',y')$-plane envelops the origin be ${p=p(\varepsilon)}$.
Obviously, $p=0$ if ${\varepsilon>\varepsilon_c}$, $p=1$ if
${\varepsilon<\varepsilon_t}$ and $0<p<1$ when
${\varepsilon_t<\varepsilon<\varepsilon_c}$.

In the case of the intermittent behavior (i.e.,
${\varepsilon_t<\varepsilon<\varepsilon_c}$) the probability of
the laminar phase with length $T$ to be observed is $P(T)=p^2$.
This period is determined by the difference of the main
frequencies of the drive ($f_d$) and response ($f_r$) systems and
may be calculated as ${T\approx 1/|f_r-f_d|}$. For the control
parameter set mentioned above, we have $T\approx80$. It is clear
that the probability of  laminar phases with length $nT$ to arise
is $P(nT)=(1-p)^{n-1}p^2$. So, the distribution of the laminar
phases with generic length $\tau$ should scale as
\begin{equation}
N(\tau)\sim p^2(1-p)^{\tau/T-1}, \mbox{\quad$\tau>T$}.
\label{eq:DistribLaw}
\end{equation}
Equation~(\ref{eq:DistribLaw}) may be also rewritten in the form
\begin{equation}
N(\tau)=A\exp(k\tau), \mbox{\quad$\tau>T$} \label{eq:ExpLaw}
\end{equation}
where $k=(1/T)\ln (1-p)$, $A$ being a normalizing coefficient.
Thus, the laminar phase distribution in the ring intermittency
 obeys an exponential law, with the parameter $k$ being
negative due to ${0<p<1}$.

\begin{figure}[tb]
\centerline{\includegraphics*[scale=0.3]{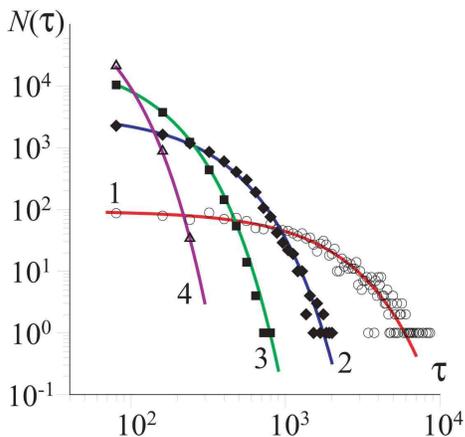}} \caption{(Color
online) The laminar phase length distributions for the different
values of the coupling strength $\varepsilon$ and exponential laws
(\ref{eq:ExpLaw}) corresponding to them. The theoretical curves are
shown by lines, the data calculated for two coupled R\"ossler
systems~(\ref{eq:Roesslers}) are shown by points. The coupling
strength values $\varepsilon$, the probability $p$ and the parameter
of the exponential law $k$ have been selected as follows: (1)
$\varepsilon=0.123$ ($\circ$), $p=0.06$, $k=-7.7\times10^{-4}$; (2)
$\varepsilon=0.12$ ($\blacklozenge$), $p=0.31$,
$k=-4.6\times10^{-3}$; (3) $\varepsilon=0.115$ ($\blacksquare$),
$p=0.65$, $k=-1.3\times10^{-2}$; (4) $\varepsilon=0.11$
($\vartriangle$), $p=0.96$, $k=-4.0\times10^{-2}$.}
\label{fgr:LengthDistrib}
\end{figure}

Let us compare the obtained theoretical law~(\ref{eq:ExpLaw}) with
the results of the numerical calculations of the intermittent
behavior of two coupled R\"ossler systems~(\ref{eq:Roesslers}). In
Fig.~\ref{fgr:LengthDistrib} the distribution of the laminar phase
lengths are shown for the different values of the coupling
strength $\varepsilon$. In the same Figure we report also the
exponential fits obeying the law~(\ref{eq:ExpLaw}). The value of
the probability $p(\varepsilon)$ to calculate the coefficient $k$
may be estimated as follows. If the length of the time realization
under consideration is $L$ (in our calculations the length of the
analyzed time series was $L=2\times10^{6}$ time units) and $N$
phase slips have been observed during this period, then the
probability $p(\varepsilon)$ that the phase trajectory on the
rotating $(x',y')$-plane envelops the origin is $p=NT/L$. E.g.,
when the coupling strength has been fixed as $\varepsilon=0.12$
(see curve~2 and points $\blacklozenge$ in
Fig.~\ref{fgr:LengthDistrib} for the distribution of the laminar
phase lengths) we have observed $N=7,772$ phase slips during the
analyzed time series, and therefore the probability
$p(\varepsilon)$ was estimated as $p=0.31$. From
Fig.~\ref{fgr:LengthDistrib} one can see an excellent agreement of
the numerical data with the theoretical law~(\ref{eq:ExpLaw}) for
the whole range of coupling strength values
${\varepsilon_t<\varepsilon<\varepsilon_c}$ where the ring
intermittency takes place.

\begin{figure}[tb]
\centerline{\includegraphics*[scale=0.3]{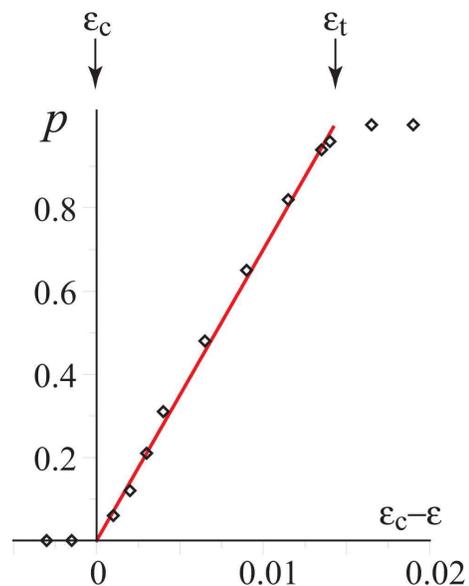}} \caption{(Color
online) The dependence of the probability $p$ that the phase
trajectory on the rotating $(x',y')$-plane envelops the origin on
the deviation of the coupling strength from the critical value
${(\varepsilon_c-\varepsilon)}$. The points obtained numerically has
been shown by symbols $\lozenge$, the linear approximation
$p=a(\varepsilon_c-\varepsilon)$ (where $a=70$) is shown by a solid
line. The critical values $\varepsilon_c$ and $\varepsilon_t$ of the
coupling parameter are shown by arrows. Note,
equation~(\ref{eq:PvsVarepsilon}) is applicable only in the coupling
strength range ${\varepsilon_t<\varepsilon<\varepsilon_c}$.}
\label{fgr:PvsVearepsilon}
\end{figure}

Let us now derive the dependence of the mean length
$\langle\tau\rangle$ of the laminar phases (i.e., the averaged
time interval between two successive phase slips) on the coupling
strength $\varepsilon$. From equation~(\ref{eq:ExpLaw}) one can
easily obtain the relationship between the mean length
$\langle\tau(\varepsilon)\rangle$  and the probability
$p=p(\varepsilon)$
\begin{equation}
\langle\tau\rangle=T-\frac{1}{k}=T-\frac{T}{\ln(1-p)}.
\label{eq:MeanLength}
\end{equation}
We have numerically observed that, in the coupling strength range
${\varepsilon_t<\varepsilon<\varepsilon_c}$ where the ring
intermittent behavior is observed, the value of the probability
$p$ is directly proportional to the deviation of the coupling
strength $\varepsilon$ from the critical value $\varepsilon_c$,
i.e.,
\begin{equation}
p(\varepsilon)\sim \left(\varepsilon_c-\varepsilon\right).
\label{eq:PvsVarepsilon}
\end{equation}
Indeed, in Fig.~\ref{fgr:PvsVearepsilon} the dependence of the
probability $p$ on the deviation of the coupling strength from the
critical value ${(\varepsilon_c-\varepsilon)}$ is shown. The
probability $p(\varepsilon)$ for each value of  $\varepsilon$ has
been calculated in the same way as it was described above.

\begin{figure}[tb]
\centerline{\includegraphics*[scale=0.3]{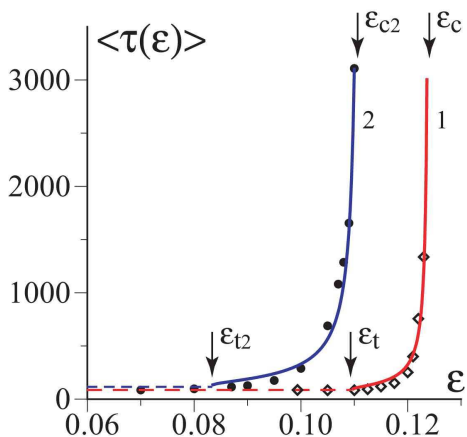}} \caption{(Color
online) The mean length $\langle\tau\rangle$ of the laminar phases
{\it vs.} $\varepsilon$ calculated numerically ($\lozenge$) for two
coupled R\"ossler systems~(\ref{eq:Roesslers}), the theoretical
curve~(\ref{eq:MeanLength2}) (red solid line 1, applicable only in
the range $\varepsilon_{t}<\varepsilon<\varepsilon_{c}$ shown by
arrows), and the asymptotic ($\varepsilon<\varepsilon_{t}$) value
of~(\ref{eq:MeanLength})  (red dashed line). The analogous curves
(filled dots, blue solid line, blue dashed line) refers to the case
of two coupled generators with tunnel diodes of Ref.
\cite{Rosenblum:1997_SynhroHuman}. In this latter case, arrows
point to $\varepsilon_{t2}=0.083$ and $\varepsilon_{c2}=0.111$,
delimiting the validity range of Eq.~(\ref{eq:MeanLength2}). }
\label{fgr:MeanLength}
\end{figure}

Since the probability $p$ on the coupling strength in the range
${\varepsilon_t<\varepsilon<\varepsilon_c}$ relates to the
coupling strength as
${p(\varepsilon)=(\varepsilon_c-\varepsilon)/(\varepsilon_c-\varepsilon_t)}$,
one easily obtains that  the dependence of the mean laminar phase
length on $\varepsilon$ has to scale in the form
\begin{equation}
\langle\tau(\varepsilon)\rangle=T\left(1-\ln^{-1}\left(\frac{\varepsilon-\varepsilon_t}{\varepsilon_c-\varepsilon_t}\right)\right).
\label{eq:MeanLength2}
\end{equation}

Notice that Eq. (\ref{eq:MeanLength2}) describes scaling
properties for the laminar periods during the ring intermittency
that are completely different from those typical of type-I
intermittency characterizing the transition to phase locking of
periodic oscillators in the presence of noise
\cite{Boccaletti:2002_SynchroPhysReport}, as well as from those
arising from the super-long laminar behavior (or eyelet
intermittency) characterizing pre-transitional stages of slightly
mismatched chaotic oscillators
~\cite{Pikovsky:1997_EyeletIntermitt,Boccaletti:2002_LaserPSTransition_PRL}.
Obviously, equation~(\ref{eq:MeanLength2}) is correct only in the
coupling strength range
${\varepsilon_t<\varepsilon<\varepsilon_c}$, whereas
(\ref{eq:MeanLength}) may be used both below $\varepsilon_t$ and
above $\varepsilon_c$. The mean length $\langle\tau\rangle$ of the
laminar phases is about $T$ for values of $\varepsilon$ is below
the critical value $\varepsilon_t$ (the asynchronous regime where
the phase slips follow one another at approximately equal time
intervals), and is  infinity in the case
$\varepsilon>\varepsilon_c$. Note also, that
$\lim_{\varepsilon\rightarrow+\varepsilon_t}\langle\tau(\varepsilon)\rangle=T$
and
$\lim_{\varepsilon\rightarrow-\varepsilon_c}\langle\tau(\varepsilon)\rangle=+\infty$
in perfect agreement with the system's behavior at the two
boundaries of the intermittency phenomenon
${(\varepsilon_t;\varepsilon_c)}$.

Finally, in Fig.~\ref{fgr:MeanLength} the dependence of the mean
laminar phase length $\langle\tau\rangle$ on the coupling strength
$\varepsilon$ is shown. The points ($\lozenge$) correspond to the
calculated mean lengths of the laminar phases and the red solid
line~1 reports the theoretical equation~(\ref{eq:MeanLength2}).
Again, one can see a perfect agreement of the theoretical curve
with the calculated points. Below the threshold $\varepsilon_t$,
the calculated mean length of the laminar phases is compared with
the asymptotic value $\langle\tau\rangle=T$ (red dashed line). In
order to show that our analysis is not limited to the chaotic
system (\ref{eq:Roesslers}), we report in the same Figure the
analogous curves (filled dots, blue solid line, blue dashed line)
obtained for the case of two unidirectionally coupled generators
with tunnel diodes described in
Ref. \cite{Rosenblum:1997_SynhroHuman}, where
Eq.~(\ref{eq:MeanLength2}) is valid in the range
$\varepsilon_{t2}<\varepsilon<\varepsilon_{c2}$
($\varepsilon_{t2}=0.083$ and $\varepsilon_{c2}=0.111$). Finally,
we would like to stress that the same intermittent scenario can be
observed in system (1) for fixed (large enough) values of coupling
strength, when varying the parameter mismatch.

In conclusion, we have reported for the first time a new type of
intermittency behavior occurring at the onset of phase
synchronization regimes of two unidirectionally coupled chaotic
oscillators with sufficiently detuned natural frequencies. Such a
type of ring intermittency differs remarkably from all the other
types of intermittency known so far. It may be observed in a
certain range of coupling parameter strengths, where the
distribution of the laminar phase lengths obeys an exponential
law. The theoretical equation for the dependence of the mean
length of the laminar phases on the coupling strength has also
been given, and are in perfect agreement with the numerically
obtained data. Though the characterization of the new intermittent
process has been here explicitly derived at the boundary of phase
synchronization of chaotic systems, we expect that the very same
mechanism can be observed in many other relevant circumstances, as
e.g. laser systems~\cite{Boccaletti:2002_LaserPSTransition_PRL},
or in the case of the interaction between the main rhythmic
processes in the human cardiovascular
system~\cite{Prokhorov:2003_HumanSynchroPRE,
Hramov:2006_Prosachivanie}.

Work partly supported by Russian Foundation of Basic Research
(projects 05--02--16273 and 06--02--16451), and by the ``Dynasty''
Foundation. S.B. acknowledges the Yeshaya Horowitz Association
through the Center for Complexity Science.


\end{document}